\documentclass[]{spie}  %>>> use for US letter paper
 \usepackage{float} 

\usepackage{amsmath,amsfonts,amssymb}
\usepackage{graphicx}
\usepackage[colorlinks=true, allcolors=blue]{hyperref}

\title{Signal processing and data acquisition system \\for the BTO detectors onboard COSI}

\author[a*]{Shunsaku Nagasawa}
\author[b, c]{Tadayuki Takahashi}
\author[d]{Kazuhiro Nakazawa}
\author[a]{Hannah Gulick}
\author[a]{Claire Chen}
\author[e,f]{Hiroki Yoneda}
\author[d]{Keigo Okuma}
\author[a]{Matt Dexter}
\author[a]{Josh Forgione}
\author[a]{Nihal Gulati}
\author[a]{Andrew Ji}
\author[a]{Kaylie Ching}
\author[a]{Isabel Schmidtke}
\author[a]{Sarah Jauregui}
\author[a]{Eric Yang}
\author[a]{Andreas Zoglauer}
\author[a]{Juan Carlos Martinez Oliveros}
\author[a]{John Tomsick}

\affil[a]{\normalsize Space Sciences Laboratory, University of California, Berkeley, 7 Gauss Way, Berkeley, CA 94720, USA}
\affil[b]{Kavli Institute for the Physics and Mathematics of the Universe (Kavli IPMU, WPI),The University of Tokyo, 5-1-5 Kashiwanoha, Kashiwa, Chiba 277-8583, Japan}
\affil[c]{International Center for Quantum-field Measurement Systems for Studies of the Universe and Particles (QUP, WPI), KEK, Ibaraki 305-0801, Japan}
\affil[d]{Department of Physics, Nagoya University, Aichi 464-8602, Japan}
\affil[e]{The Hakubi Center for Advanced Research, Kyoto University, Sakyo, Kyoto 606-8501, Japan}
\affil[f]{Department of Physics, Kyoto University, Kitashirakawa Oiwake-cho, Sakyo-ku, Kyoto 606-8502, Japan}

\authorinfo{*Shunsaku Nagasawa: E-mail: nshunsaku@berkeley.edu}

\begin{document} 
\maketitle

\begin{abstract}
The energy range from a few hundred keV to a few MeV includes important probes such as nuclear gamma-rays and the 511 keV annihilation line. 
However, compared to X-rays and GeV/TeV gamma-rays, this range suffers from lower sensitivity by orders of magnitude. The upcoming NASA SMEX satellite mission Compton Spectrometer and Imager (COSI), scheduled for launch in 2027, is expected to break through this limitation with its Compton telescope utilizing a germanium semiconductor detector, covering the 0.2-5 MeV energy range.
In addition to the main instrument, two Background and Transient Observer (BTO) detectors will be installed on COSI. The detectors are NaI(Tl) scintillators coupled with SiPMs, and they are being developed as a student collaboration project. BTO aims to 1) measure background radiation in orbit to maximize COSI’s sensitivity and 2) detect GRBs and other gamma-ray transients.
For this purpose, it is required to cover the lower-energy range from 30 keV to 2 MeV with $<$20\% FWHM energy resolution. In addition, large signals and afterglow generated by heavy ions penetrating the NaI(Tl) crystal should be appropriately handled.
To address these requirements, we have developed a compact signal processing and data acquisition system comprised of two main components: an analog board and a digital board. 
The analog board amplifies signals from the SiPM, generates triggers, and performs AD conversion. The digital board features a Microchip SAMV71 microcontroller, and we developed the software to control the analog board, read ADC data via SPI interface, add timestamps, and buffer event data. Through this development, we achieved the required wide bandpass and an energy resolution of 10\% FWHM at 662 keV with a processing time of 20 $\mathrm{\mu}$s per event. We also implemented veto signal generation for large signals using a discriminator and an onboard detection algorithm for transient events.
\end{abstract}


\section{Introduction} \label{sec:intro}
The energy range from a few hundred keV to several MeV serves as a critical probe of diverse high-energy astrophysical phenomena \cite{de2021gamma}. It bridges the range between thermal emission at keV energies and non-thermal processes at GeV and higher energies. This band also provides key probes such as the 511 keV electron-positron annihilation line\cite{kierans2020detection, siegert2023positron}, nuclear gamma-ray lines from radioactive elements\cite{beechert2022measurement, liu2025gamma}, and potential signals from dark matter particle interactions\cite{bartels2017prospects, caputo2023dark, watanabe2025light}. 
However, this band remains one of the least explored regimes in multi-wavelength astrophysics, with instrumental sensitivity orders of magnitude lower than in neighboring X-ray and GeV/TeV gamma-ray bands. This long-standing limitation is commonly referred to as the ``MeV gap"\cite{takahashi2012astro, kierans2022compton}.

The upcoming NASA SMEX satellite mission Compton Spectrometer and Imager (COSI)\cite{Tomsick2024}, scheduled for launch in 2027, aims to address this gap. COSI employs a Compton telescope covering the 0.2 to 5 MeV energy range, based on high-purity germanium semiconductor detectors\cite{beechert2022calibrations}. 
It provides imaging, spectroscopy, and polarimetry of gamma-ray sources and surveys the entire sky with a wide field of view (FOV) of $>$25\% of sky.
The excellent energy resolution of the germanium detectors enables high line sensitivity, making COSI particularly well suited for studies of nuclear gamma-ray lines and positron annihilation.

To fully exploit the capabilities of COSI and achieve high-sensitivity observations, continuous monitoring and accurate modeling of the in-orbit gamma-ray background is essential. 
Instruments operating in the MeV range are particularly affected by background components such as atmospheric albedo radiation and activation-induced gamma-ray lines from cosmic-ray interactions\cite{tatischeff2024orbits, kierans2022compton, schonfelder2004lessons}.
These background sources significantly contribute to the detector count rate and must be carefully characterized and modeled to ensure accurate interpretation of astrophysical gamma-ray signals.
In addition, the capability to detect transient gamma-ray phenomena such as gamma-ray bursts (GRBs) and magnetar flares is essential for advancing time-domain and multi-messenger astrophysics. 
Short gamma-ray bursts are predominantly associated with the coalescence of compact binary systems, such as neutron star–neutron star and possibly neutron star–black hole mergers\cite{d2015short}. The detection of gamma-ray counterparts to gravitational wave events supports time-resolved multi-messenger studies, offering complementary insights into the underlying emission mechanisms.  In addition, a subset of short GRBs originates from giant flares of soft gamma-ray repeaters (SGRs)\cite{ofek2008grb,frederiks2007possibility,yang2020grb}, which are believed to arise from highly magnetized neutron stars called magnetars. The relative contribution of such magnetar-driven flares to the overall short GRB population is not well-constrained.  Comprehensive characterization of these events requires broadband temporal and spectral measurements with sufficient sensitivity.

To fulfill these two primary objectives, background characterization and transient detection, COSI will have two independent gamma-ray detectors called Background and Transient Observer (BTO)\cite{gulick2024across,gulick2025study}. 
BTO is a student collaboration project, and undergraduate students at UC Berkeley play a central role in its development.
BTO extends the COSI bandpass to lower energies by providing spectral coverage in the 30 keV to 2 MeV range.
With the help of the overlapping energy coverage and shared field of view between COSI and BTO, the combined system enables broadband spectroscopic observations of gamma-ray transients.
Moreover, the BTO detector has a simple and well-understood structure, consisting of a NaI(Tl) scintillator coupled with a silicon photomultiplier (SiPM), which facilitates accurate modeling of the background from measured data.

In this paper, we present the development of the signal processing and data acquisition system for the BTO detectors onboard COSI.
The readout system is required to cover a wide energy bandpass from 30 keV to 2 MeV with an energy resolution better than 20\% FWHM.
In addition, heavy ion interactions in the NaI(Tl) crystal can produce large-amplitude signals followed by afterglow\cite{gulick2025study}, which must be properly identified and rejected.
To meet these requirements, we designed a compact data acquisition system composed of an analog board and a digital board.
Section \ref{sec:bto_overview} provides an overview of the BTO detectors and electronics systems.
Sections \ref{sec:analog} and \ref{sec:digital} describe the details of analog and digital signal processing, including the software architecture and onboard transient event detection algorithms.
In Section \ref{sec:performance}, we present results from a spectroscopic performance evaluation using a prototype BTO detector and readout system, demonstrating that the developed system meets the performance requirements for BTO.

\section{The Background and Transient Observer}\label{sec:bto_overview}
\subsection{BTO Detectors} \label{sec:bto_detector}
\begin{figure}[H]
    \centering
    \includegraphics[width=1.0\linewidth]{ 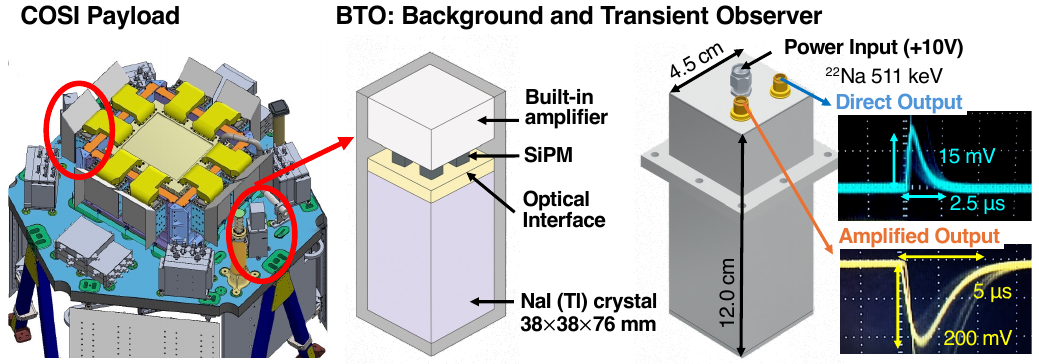}
    \caption{Left: Schematic diagram of the COSI Payload Interface Plate (PIP). The two BTO detectors are mounted on opposite sides to enable coarse transient localization. Right: A schematic diagram of the Scionix detector module, featuring a NaI(Tl) scintillator with SiPMs. The detector has two outputs: a direct output used for rejecting large signals caused by heavy ion interactions (blue waveform), and an amplified output used for energy measurements (yellow waveform).
    The direct and amplified output waveforms, corresponding to 511 keV gamma-rays from a \textsuperscript{22}Na source, are shown with 50 $\Omega$ and 1 M$\Omega$ termination, respectively.
}
    \label{fig:fig_bto_det}
\end{figure}

The BTO detector is based on Scionix detector modules and consists of a NaI(Tl) scintillator (38 mm$\times$38 mm$\times$76 mm) coupled to three ArrayJ-60035-4P-PCB SiPMs (see Figure \ref{fig:fig_bto_det} right). The scintillator is enclosed in a 4.5 cm$\times$4.5 cm$\times$12 cm aluminum housing, with a single SMA input for +10V power and two SMA signal outputs: the ``direct output" and the ``amplified output". 
The direct output is taken directly from the SiPMs in the detector while the amplified output is processed by the built-in preamplifier in the Scionix detector module. 
The direct output has a pulse rise time (0-100\%) of 370 ns and a fall time (1/e) of 900 ns, with a gain of approximately 27 mV/MeV when terminated with 50 $\Omega$ resistance.
The amplified output has a pulse rise time is 1.3 $\mathrm{\mu s}$, and a fall time is 2.1 $\mathrm{\mu s}$ with a gain of approximately 400 mV/MeV when terminated with 1 M$\Omega$ resistance. 
The amplified output is used primarily for energy measurements due to its higher gain. In contrast, the direct output offers a larger dynamic range and can be used to identify large signals caused by heavy ion interactions and enables the rejection of false triggers from afterglows \cite{gulick2025study}.

The two BTO detectors are mounted on the opposite sides of the COSI Payload Interface Plate (PIP) (see Figure \ref{fig:fig_bto_det} left), optimizing their FOV to cover more than 60\% of the sky.
This geometry also enables the use of the shadow cast by the germanium detectors, BGO shields, and other structures within the COSI payload to provide coarse localization of events occurring outside the nominal COSI FOV, distinguishing between solar, terrestrial, and astrophysical origins \cite{gulick2024across}.
To avoid obstruction by surrounding components, the mechanical housing for each BTO detector includes a pedestal structure that raises the scintillator crystal by 11.5 cm. This design ensures that the full COSI FOV is covered by a portion of each BTO's FOV. Consequently, any transient event detected by a BTO within this overlapping region can be localized to within a few degrees using the COSI germanium detectors.

\subsection{BTO Electronics Overview}\label{sec:bto_electronics_overview}
\begin{figure}[h]
    \centering
    \includegraphics[width=1\linewidth]{ 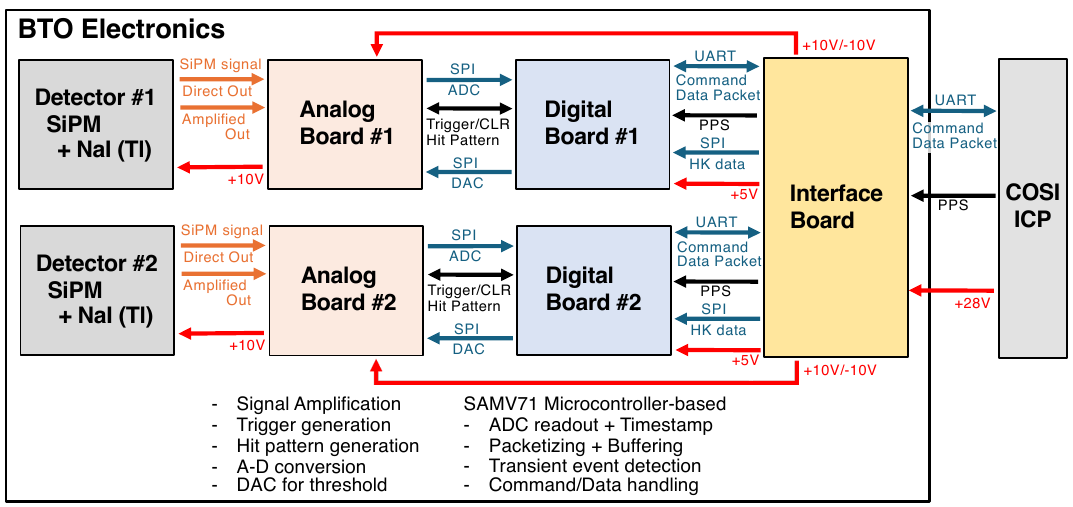}
    \caption{Overview of the BTO electronics architecture. Each detector provides two outputs: a direct output for large-signal discrimination and an amplified output for energy measurements. The signals are processed by dedicated analog and digital boards. The analog boards perform amplification, triggering, and ADC, while the digital boards, based on SAMV71 microcontrollers, handle event processing, packet generation, and transient event detection. The digital boards communicate with the COSI ICP via a common interface board, which also supplies power and distributes the PPS signal for time synchronization.}
    \label{fig:fig_bto_electro_overview}
\end{figure}

Figure \ref{fig:fig_bto_electro_overview} shows the overview of the BTO electronics architecture.
The signal from each detector is processed independently by a dedicated analog board and a digital board. All digital boards communicate with the COSI Instrument Control Package (ICP) through a common interface board, which handles the bidirectional exchange of command and data packets. 

The analog board (see Section~\ref{sec:analog}) receives the amplified output signal from the detector and performs signal amplification, trigger generation, and analog-to-digital conversion (ADC). In addition, based on the various discriminators, the analog board can set trigger thresholds and generate hit patterns using both the amplified and direct output signals.
The digital board (see Section \ref{sec:digital}), based on the Microchip SAMV71 microcontroller, manages the operation of the analog board and handles command communication.  In response to trigger signals from the analog board, it reads out ADC data, adds timestamps, builds data packets, and transfers them to the ICP on request. It also implements an onboard transient detection algorithm (see Section~\ref{sec:detect_grb}) that switches measurement modes in response to detected events.
The interface board forwards commands from the ICP to each digital board and transmits data packets from the digital boards back to the ICP. It also generates and distributes the power required for the analog and digital boards from a +28 V supply provided by the ICP. Furthermore, the interface board hosts temperature sensors and voltage/current monitors for each detector and board. These housekeeping data are obtained by the digital board and sent to the ICP. The interface board also 
distributes the Pulse-Per-Second (PPS) signal from the ICP to each digital board for time synchronization.

\begin{figure}[h]
    \centering
    \includegraphics[width=1\linewidth]{ 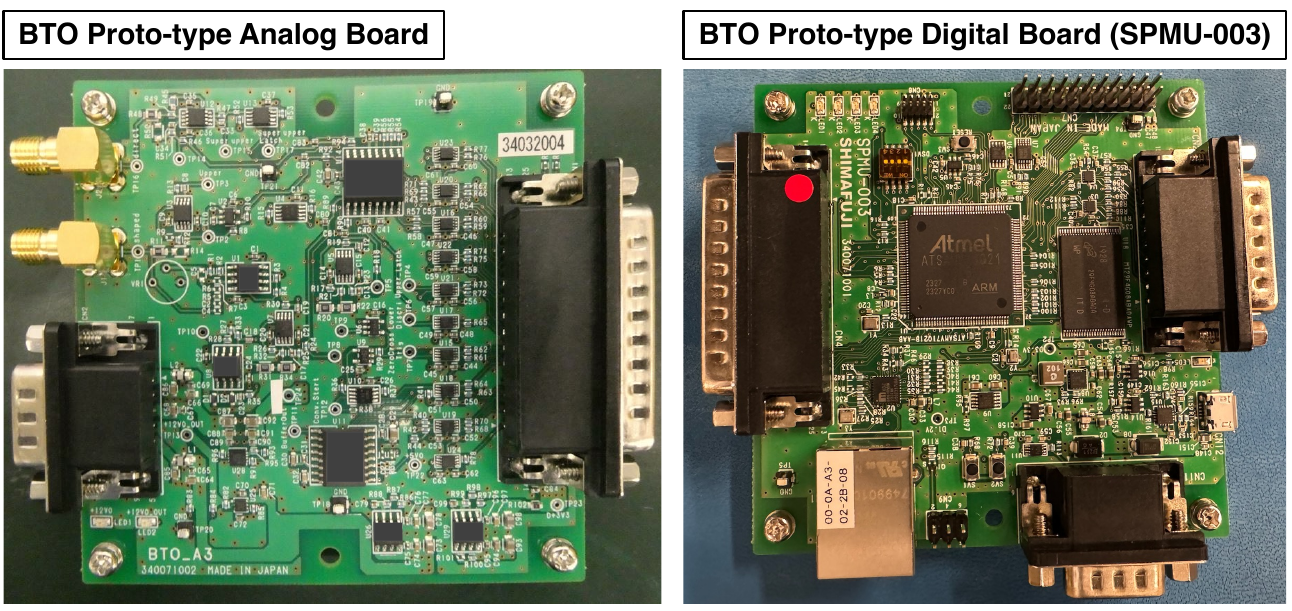}
    \caption{Photos of prototype analog and digital boards (SPMU-003).  The boards are designed to fit within an area of 90 mm$\times$90 mm. While standard D-sub connectors are used in the prototype, they will be replaced with micro D-sub connectors in the flight version.}
    \label{fig:fig_bto_proto_pic}
\end{figure}

Figure \ref{fig:fig_bto_proto_pic} shows pictures of prototype analog and digital boards. Both boards are designed to fit within an area of 90 mm$\times$90 mm.
The pin configuration is identical to that intended for the flight model. Although standard D-sub connectors are used in the prototype, they will be replaced with micro D-sub connectors in the flight model.
The prototype digital board, also referred to as SPMU-003, is a general-purpose microcontroller board co-developed by Shimafuji Electric Incorporated (\url{http://www.shimafuji.co.jp/en/}) and Kavli IPMU, The University of Tokyo. 

\section{Analog signal processing}\label{sec:analog}
\begin{figure}[h]
    \centering
    \includegraphics[width=1.\linewidth]{ 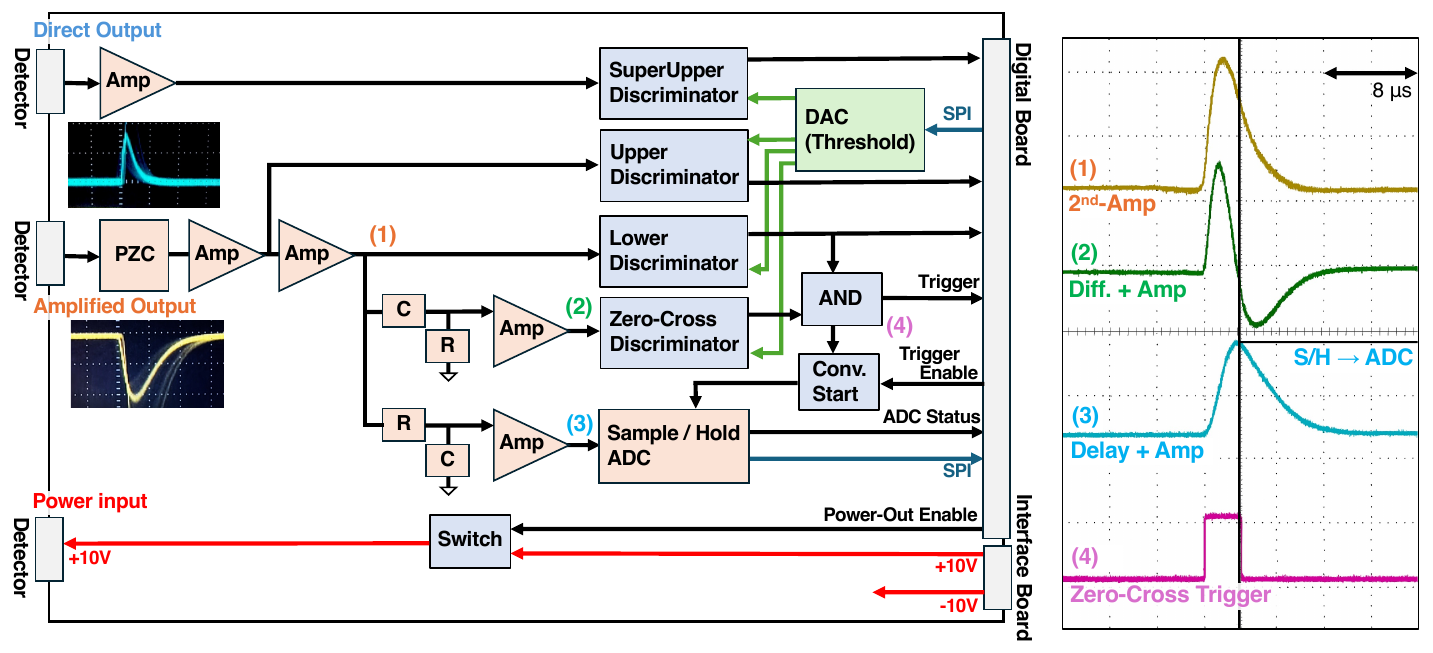}
    \caption{Schematic diagram of the signal processing flow implemented on the BTO analog board, along with representative waveforms at each processing stage. The yellow waveform shows the amplified signal, the green waveform is its differentiated bipolar form used for triggering, the blue waveform is the delayed signal used for peak sampling, and the magenta pulse indicates the trigger timing.}
    \label{fig:fig_bto_analog_board}
\end{figure}

Figure \ref{fig:fig_bto_analog_board} shows a schematic of the signal processing flow implemented on the analog board. The direct output from the BTO detector is processed by a ``SuperUpper discriminator", which identifies large signals of $\gtrsim$ 10 MeV, typically caused by heavy ion interactions.
The timing of these signals is latched and recorded by the digital board, and is used later to identify and reject events caused by afterglow.

Energy measurement is performed using the amplified output signal. The corresponding signal waveforms at each processing stage are shown in the right panel of Figure \ref{fig:fig_bto_analog_board}.
First, the signal passes through a pole-zero cancellation (PZC) circuit and is then amplified (yellow waveform). The PZC time constant was optimized based on beamline experiments with heavy ions\cite{gulick2025study}, ensuring rapid baseline recovery following large energy depositions.
``Upper discriminator" is also applied to detect large gamma-ray signals of $\gtrsim$ 2 MeV and events with saturated signals are flagged accordingly.
The amplified signal is then split into two paths. In the first path, the signal is shaped into a bipolar waveform via a differentiator circuit (green waveform), and a trigger is generated when both the ``Lower discriminator" and ``Zero-Cross discriminator" conditions are satisfied, which means that the pulse height exceeds a certain threshold and the signal reaches its zero-crossing point.
In the second path, the signal is delayed using a delay circuit with the same time constant as the differentiator (blue waveform). This ensures that the peak of the delayed waveform aligns precisely with the zero-crossing point of the bipolar signal. The trigger timing determined by the zero-crossing (magenta waveform) is then used to sample and hold the peak of the delayed waveform. Finally, this peak value is digitized via ADC for energy measurement.

The threshold levels for each of the four discriminators can be independently configured via SPI interface by setting the corresponding DAC values from the digital board. Timing and counter information from the discriminators is also obtained by the digital board for further processing.
Once the ADC conversion is completed, this timing is latched by the digital board, and the digitized signal is read out via SPI interface (see Section \ref{sec:digital} for details). The analog board also includes a switch circuit that controls the +10 V power supply to the detector. This switch is toggled based on control signals received from the digital board.

\section{Digital signal processing and software}\label{sec:digital}
\subsection{Overview of BTO Digital Board}\label{sec:digital_overview}
\begin{figure}[h]
    \centering
    \includegraphics[width=0.9\linewidth]{ 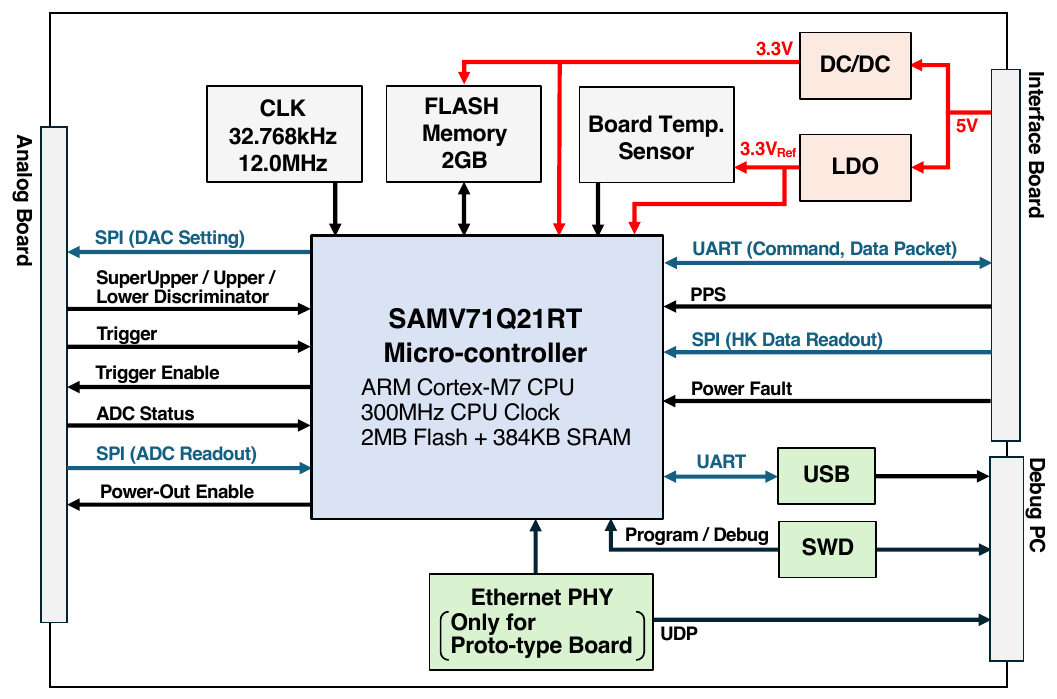}
    \caption{Block diagram of the BTO digital board. The board is based on the SAMV71 microcontroller and includes external 12 MHz and 32.768 kHz clocks for system and RTC operation, respectively. A DC-DC converter regulates 5 V input power from the interface board to 3.3 V. SWD, USB UART, and Ethernet (prototype only) interfaces are available for programming, debugging for ground tests.}
    \label{fig:fig_bto_digital_board}
\end{figure}

Figure \ref{fig:fig_bto_digital_board} shows a block diagram of the BTO digital board. 
The core of the digital board is the Microchip SAMV71 microcontroller, a 32-bit ARM Cortex-M7 processor which can operate at up to 300 MHz.
For the flight digital board, a radiation-tolerant version of this device (SAMV71Q21RT) is employed\cite{microchip_samv71q21rt}.
The SAMV71 features 16 KB of instruction cache (ICache), 16 KB of data cache (DCache), 2048 KB of embedded flash memory, and 384 KB of SRAM. Since the internal memory is limited, the digital board is equipped with 16 Gbit (2 GB) of non-volatile NAND flash memory for storing science data obtained by the microcontroller.

Since the SAMV71 operates with 3.3 V logic, the 5 V input power supplied from the interface board is regulated to 3.3 V by a low-noise DC-DC switching converter mounted on the digital board.
In addition, level shifters are used to translate 5 V digital output signals from the analog board to 3.3 V compatible levels.
External oscillator clocks of 12 MHz and 32.768 kHz are employed, with the former used in conjunction with the on-chip PLL to generate a 300 MHz system clock, and the latter serving as the Slow Clock (SLCK) for real-time clock (RTC) operation.
The digital board includes both a Serial Wire Debug (SWD) and a USB UART interface for on-ground programming and debugging. These interfaces will be covered and disabled in the flight configuration.
To accommodate high-throughput testing during on-ground tests, the prototype version of the digital board (SPMU-003) is equipped with an Ethernet interface, supporting high-speed data transmission via UDP.

\subsection{Software Overview}\label{sec:digital_soft}
The software for the BTO digital board is implemented on the Microchip SAMV71 microcontroller and developed based on the Atmel Software Framework (ASF4)\cite{microchip_asf4}, a free and open-source library that provides hardware abstraction and middleware components to simplify microcontroller programming and reduce development time. 
By adopting a microcontroller-based architecture, the system allows development primarily in the C programming language using a dedicated integrated development environment, lowering the barrier to software implementation compared to FPGA-based designs. This approach is particularly well suited for student-led projects such as BTO.

The BTO digital board reads photon ADC data via the SPI interface from the analog board for each detected event. 
It primarily operates in a ``binned mode", in which incoming events are accumulated into histograms with a coarse energy binning over one-second intervals.
In addition, the count values of each discriminator (SuperUpper, Upper, Lower, and Zero-cross) are recorded at 50 ms interval.
However, upon detection of a transient event (as described in Section \ref{sec:detect_grb}), the operation mode is switched to an ``event-by-event" mode. 
In this mode, while continuing to record binned data, the system also logs the energy and timestamp of each individual photon, enabling higher energy and temporal resolution during gamma-ray transients. 
This mode-switching algorithm allows BTO to preserve fine time resolution during transient events while maintaining efficient use of limited telemetry bandwidth. 
Event-by-event photon data are temporarily buffered in external NAND flash memory and transmitted progressively as bandwidth permits.
Communication with the COSI ICP is conducted in ``Pull" mode, in which the ICP issues explicit requests to retrieve data packets via the interface board over the UART interface. The ICP also periodically requests housekeeping data, including detector temperature, voltage, and current, which are obtained from the interface board via SPI interface.
A software watchdog timer (WDT) is implemented to monitor system health and automatically reset the microcontroller in case of unexpected software hangs or failures.

\begin{figure}[h]
    \centering
    \includegraphics[width=1\linewidth]{ 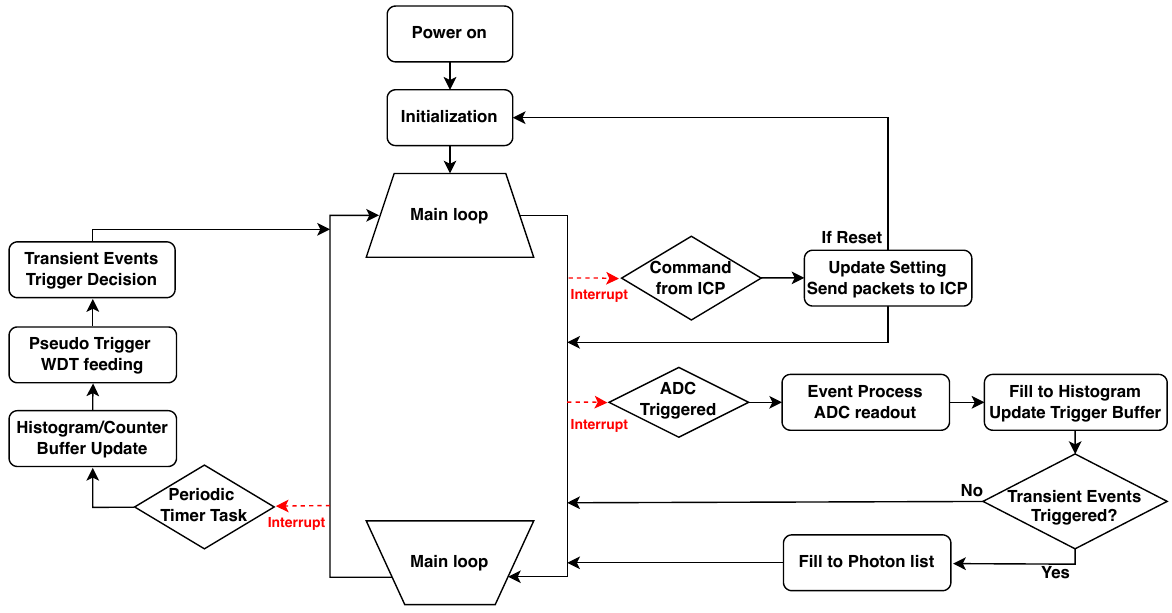}
    \caption{Overview of the BTO digital board software. The system runs a main loop with three key interrupt routines: a Periodic Timer Task Interrupt for buffer updates and transient detection, an ADC Trigger Interrupt for event readout from the analog board, and a Command Interrupt for command reception and data transmission via UART from the interface board.}
    \label{fig:fig_bto_software_diagram}
\end{figure}

Figure \ref{fig:fig_bto_software_diagram} shows the overview of the software concept.
In this architecture, the microcontroller runs mainly on a main loop, while essential functions such as data acquisition and communication are handled through specific interrupt-driven routines.
Three primary types of interrupt routines are implemented: ``Periodic Timer Task Interrupt",  ``ADC Trigger Interrupt" and ``Command Interrupt".
The Periodic Timer Task Interrupt is driven by the Timer driver provided in ASF4 and generates interrupts at fixed intervals (e.g., every 50 ms) to execute registered tasks.
In this software, the scheduler handles operations such as updating histograms and counter buffers, feeding the software WDT, and evaluating transient event triggers based on count rates (see Section \ref{sec:detect_grb} for details).
Additionally, pseudo-random triggers are inserted into the photon list at a known rate (e.g., 2 Hz) within the timer task, and the dead time is estimated by comparing the number of inserted triggers with the number of successfully recorded ones\cite{kokubun2007orbit}.

\begin{figure}[h]
    \centering
    \includegraphics[width=1\linewidth]{ 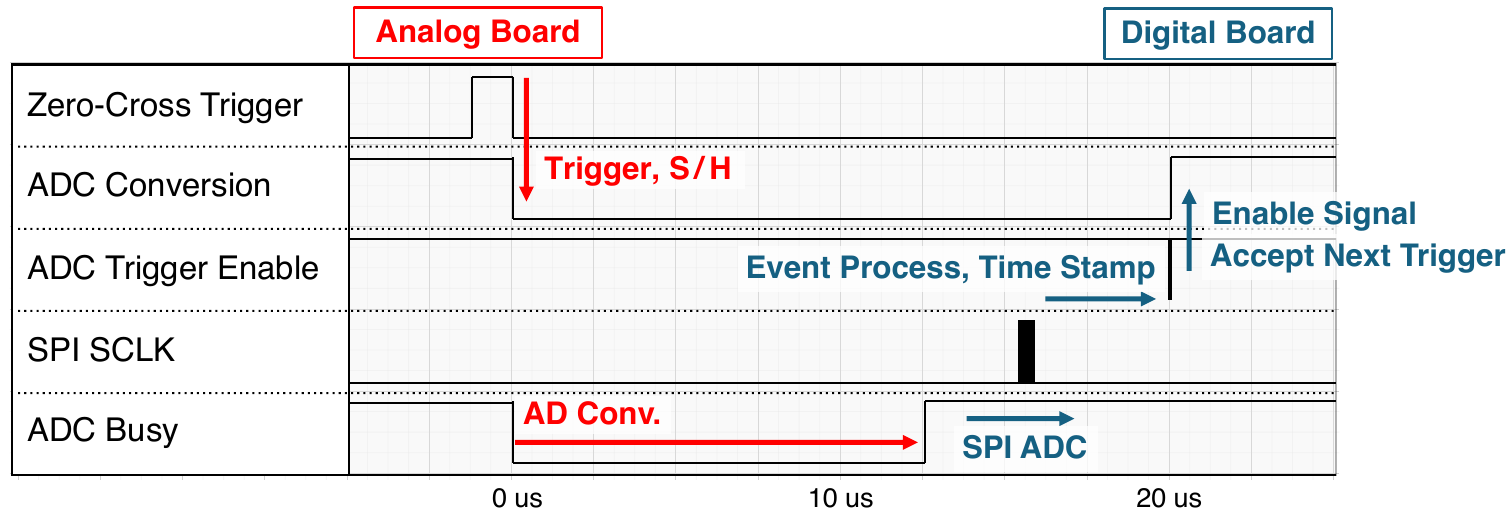}
    \caption{Timing diagram of event readout between the BTO analog and digital boards. Upon zero-crossing, the pulse height is sampled and held for ADC conversion. After the conversion completes and the BUSY signal of ADC status returns high, the digital board issues an interrupt, retrieves the data via SPI interface, processes the event, and sends an enable signal to accept the next trigger. The total dead time is approximately 20 $\mathrm{\mu}$s per event.}
    \label{fig:fig_bto_timing}
\end{figure}

The ADC Trigger Interrupt handles ADC event readout by capturing trigger signals from the analog board. The timing of relevant digital signals during readout is illustrated in Figure \ref{fig:fig_bto_timing}. On the analog board, when the differentiated waveform for triggering reaches its zero-crossing point, the pulse height is sampled and held, and ADC conversion is initiated. Once the ADC conversion is completed, the BUSY signal of the ADC status returns to high. At this time, an interrupt is issued to the digital board, which initiates event processing routines including data readout via SPI interface, timestamping, and data buffering. After processing, a Trigger Enable signal is sent back to the analog board to allow acceptance of the next trigger. The total dead time for one event cycle is approximately 20 $\mathrm{\mu}$s, allowing the system to handle event rates of up to 50,000 counts per second.

The Command Interrupt is managed by the USART driver in ASF4 and is triggered upon receiving a UART command from the interface board. Upon this interrupt, the software interprets a command and executes the corresponding instruction. Supported operations include transmission of science and housekeeping data packets, adjustment of measurement parameters and operation modes, memory access, and software resets.
All command and telemetry packets follow the CCSDS Space Packet protocol (CCSDS 133.0-B-2, June 2020)\cite{ccsds_book}, encapsulated within an Instrument Transfer Frame (ITF). The ITF format includes a synchronization word for packet boundary detection and a checksum for error detection. All BTO data handling and interpretation are conducted on the ground.


\subsection{Transient Event Detection Algorithm}\label{sec:detect_grb}
\begin{figure}[h]
    \centering
    \includegraphics[width=1\linewidth]{ 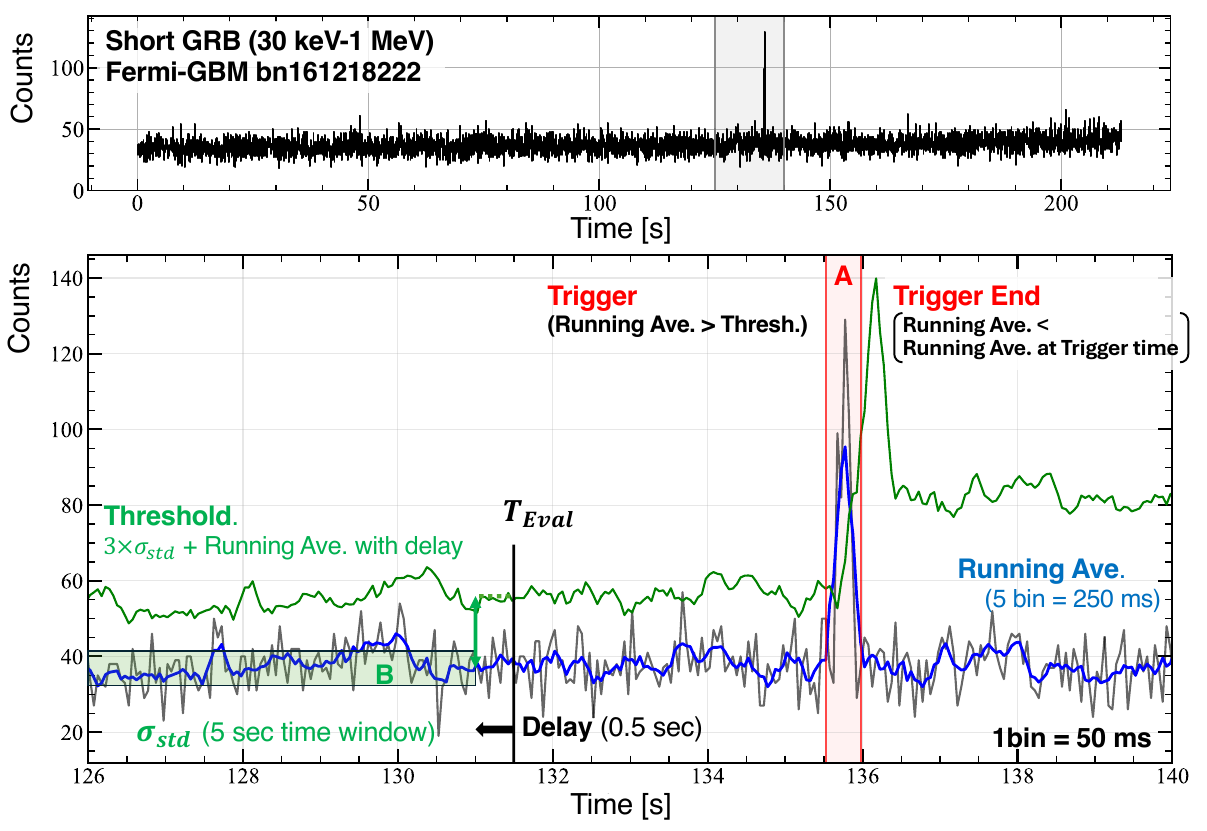}
    \caption{Schematic of the transient trigger algorithm applied to a short GRB (bn161218222) detected by Fermi-GBM. Top: Light curve of Fermi-GBM in the 30 keV–1 MeV band. Bottom: Zoom-in of the 126–140 s interval, illustrating the trigger algorithm. The 250 ms running average of count rate (blue) is compared against background fluctuations (green region B), defined as the mean plus three times the standard deviation over a preceding 5-second window.
    A trigger is issued when the average exceeds this fluctuation threshold (start of red region A) and ends once the count rate returns to the pre-trigger level (end of red region A).}
    \label{fig:fig_bto_grb_trig}
\end{figure}

To enable automatic switching of observation modes as discussed in Section \ref{sec:digital_soft}, the BTO software incorporates an onboard trigger algorithm to identify transient events, such as GRBs, and determine when to enter event-by-event mode for photon-by-photon recording.
Figure \ref{fig:fig_bto_grb_trig} illustrates the algorithm using a short GRB detected by Fermi-GBM as an example.
The software continuously monitors the background count rate at 50 ms intervals.
A transient trigger is issued when a statistically significant increase in count rate is detected (start of red region A).
Specifically, the trigger algorithm computes a running average over five bins (250 ms; shown in blue) leading up to the current time.
This is compared to the background, which is characterized by the mean and standard deviation of count rates calculated over a 5-second window ranging from 5.5 to 0.5 seconds before the current time.
If the running average exceeds the mean background level by more than three times the standard deviation (green region B), a transient trigger is issued.
This detection threshold is configurable via command.
The trigger condition is considered to have ended once the count rate returns to the pre-trigger background level (end of red region A), and the system switches back to binned mode.
To preserve sufficient context for background analysis, photon data are also recorded for a certain period before and after the trigger window.

\section{Spectroscopic Performance}\label{sec:performance}
\begin{figure}[h]
    \centering
        \includegraphics[width=1\linewidth]{ 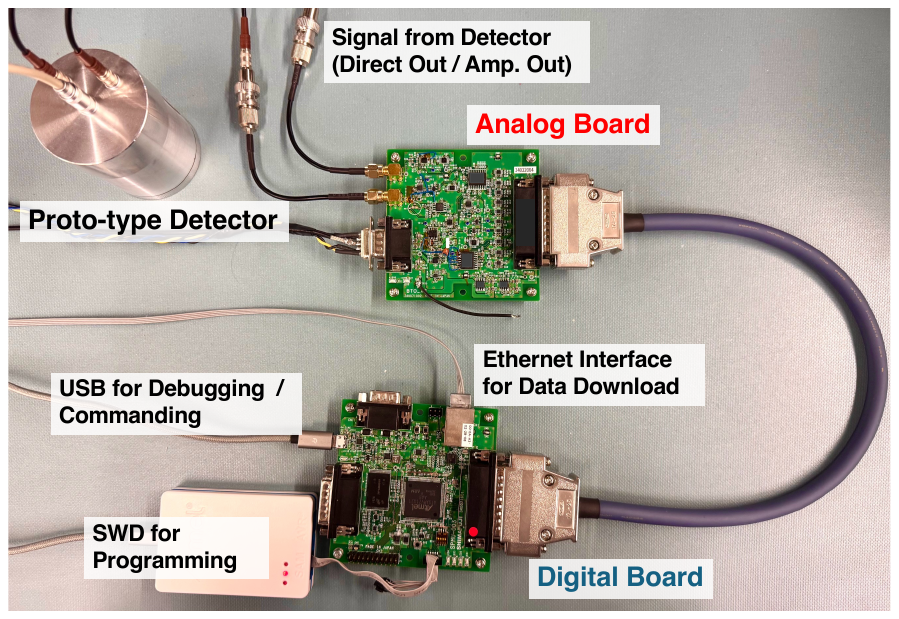}
    \caption{Test setup for spectroscopic performance evaluation using the BTO prototype system. The cylindrical BTO test detector is connected to the analog board. The proto-type analog and digital boards are connected via a D-sub 25 pin cable. The digital board is connected to a PC, featuring SWD for programming and USB UART and Ethernet interfaces for command input and data download.
    }
    \label{fig:fig_bto_proto_test_photo}
\end{figure}

We evaluated the spectroscopic performance of the prototype BTO system using the developed analog and digital boards.
Figure~\ref{fig:fig_bto_proto_test_photo} shows a setup for the performance test.
The detector used in this test is a cylindrical prototype (with a NaI(Tl) scintillator of 51 mm diameter and 51 mm height), in which only the geometry differs from the flight model; all other aspects, including signal processing and electronics, are identical to the flight configuration.
The analog and digital boards are connected via a D-sub 25-pin cable.
The digital board is connected to a PC, enabling software updates via the SWD interface, as well as command transmission and data download through the USB UART or Ethernet interfaces.
Power is supplied under the same conditions as the flight configuration: +10 V to the detector, $\pm$10 V to the analog board, and +5 V to the digital board.

\begin{figure}[h]
    \centering
    \includegraphics[width=1\linewidth]{ 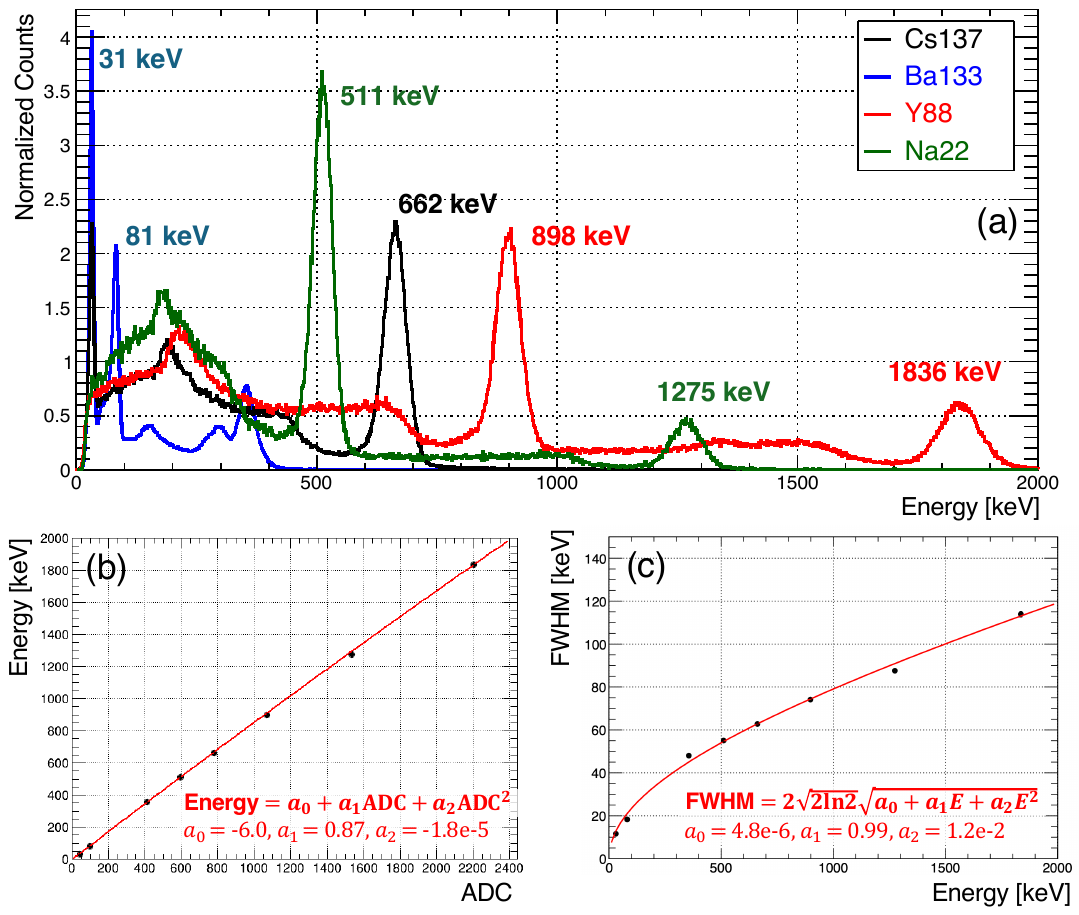}
    \caption{(a) Energy spectra obtained with the BTO prototype using $^{137}$Cs, $^{133}$Ba, $^{88}$Y, and $^{22}$Na sources.
    (b) Relationship between ADC channel and incident photon energy.
    (c) Energy dependence of the energy resolution (FWHM) for various calibration energy peaks. The data are fitted with an empirical function of $\mathrm{FWHM} = 2 \sqrt{2\ln 2} \cdot \sqrt{a_0 + a_1 E + a_2 E^2}.$.}
    \label{fig:fig_bto_proto_performance}
\end{figure}

Spectra were obtained for calibration sources including $^{137}$Cs, $^{133}$Ba, $^{88}$Y, and $^{22}$Na. The energy spectra obtained are shown in Figure \ref{fig:fig_bto_proto_performance}(a). 
These results demonstrate that the BTO successfully covers the required broad energy range from 30 keV to 2 MeV. Figure \ref{fig:fig_bto_proto_performance}(b) shows the relationship between ADC channels and incident photon energy, confirming good linearity across the full dynamic range.
Figure \ref{fig:fig_bto_proto_performance}(c) shows the energy resolution (FWHM) as a function of photon energy for various calibration peaks. The dependence is well described by the common empirical function: $\mathrm{FWHM} = 2 \sqrt{2\ln 2} \cdot \sqrt{a_0 + a_1 E + a_2 E^2}.$, which reflects contributions from electronic noise, statistical fluctuations, and charge other non-linearity effects.
At the 662 keV peak from $^{137}$Cs, an energy resolution of 65 keV (FWHM) was achieved, satisfying the required performance of $<$20\% resolution.


\section{Conclusion}\label{sec:conclusion}
We have developed the Background and Transient Observer (BTO) as an independent gamma-ray instrument for the COSI mission, with two main objectives: in-orbit characterization of the background and detection of transient gamma-ray events such as GRBs and magnetar flares. These capabilities are critical for enhancing the sensitivity of the primary Compton telescope and enabling time-resolved multi-messenger studies.

The BTO detector consists of a NaI(Tl) scintillator coupled to a SiPM (Section \ref{sec:bto_detector}). To read out and process the signals from the detector, we designed a compact data acquisition system composed of an analog board and a digital board (Section \ref{sec:bto_electronics_overview}).
The analog board amplifies the detector signals, generates trigger signals, and performs analog-to-digital conversion (Section \ref{sec:analog}). 
It utilizes two distinct signal outputs with different dynamic ranges derived from the BTO detector: the ``amplified output," used for energy measurement in the 30 keV to 2 MeV range, and the ``direct output," dedicated to identifying large signals above approximately 10 MeV, typically caused by heavy ion interactions. The latter is processed by a ``SuperUpper discriminator" on the analog board, and its timing is latched by the digital board for later use in identifying and rejecting afterglow-related events during offline analysis.
The digital board, based on a Microchip SAMV71 microcontroller, handles event processing including SPI-based ADC readout, timestamping, data buffering, and command communication with ICP via the interface board. Its software architecture combines a main loop with three key interrupt-driven routines for periodic tasks, ADC-triggered event acquisition, and command handling (Section \ref{sec:digital_overview} and \ref{sec:digital_soft}).
The dead time per event was measured to be approximately 20 $\mathrm{\mu}$s, supporting operation at rates up to 50,000 counts per second.
In addition, the digital board implements an onboard trigger algorithm to detect transient events (Section \ref{sec:detect_grb}). By comparing the running average of the count rate with its background fluctuations, the system autonomously identifies significant count enhancements. 
Upon triggering, the system switches the observation mode from binned mode, which accumulates histogram and count rates over coarse energy and time intervals, to event-by-event mode, which continues to collect these binned data while additionally recording detailed information for each detected photon.
This dynamic mode switching allows efficient use of limited telemetry bandwidth while ensuring high-resolution data recording during transient events.

To validate the performance of the BTO system, we conducted a spectroscopic evaluation using a prototype detector and electronics (Section \ref{sec:performance}) with calibration sources.
The energy spectra obtained confirmed the required bandpass from 30 keV to 2 MeV, with a sufficient linearity and the energy resolution of 65 keV FWHM at 662 keV peak.
These results demonstrate that the developed BTO system fulfills the technical requirements for background monitoring and transient detection in the COSI mission and is ready for integration into the flight model.

\acknowledgments % equivalent to \section*{ACKNOWLEDGMENTS}       
The Compton Spectrometer and Imager is a NASA Explorer project led by the University of California, Berkeley.  COSI and BTO are supported by NASA under contract 80GSFC21C0059.  
This work was supported by JSPS, Japan KAKENHI Grant Number 22J12583.
S.N. is supported by FoPM (WISE Program) and JSR Fellowship, The University of Tokyo, Japan, and by the JSPS Overseas Research Fellowship.
The authors would like to thank Jonathan Ji and Chris Moeckel for the contribution of the BTO software development.
% References
\bibliography{main} % bibliography data in report.bib
\bibliographystyle{spiebib} % makes bibtex use spiebib.bst

\end{document}